\documentclass[aip,pof,reprint,onecolumn]{revtex4-1} 
\usepackage{graphicx}
\usepackage{dcolumn}
\usepackage{bm}
\usepackage{amssymb}
\usepackage{amsmath}
\usepackage{amsfonts}
\usepackage{mathtools}
\usepackage{comment}
\usepackage{lineno}
\usepackage{soul}
\usepackage{natbib}
\usepackage{color}     
\usepackage{soul}
\usepackage[usenames,dvipsnames,svgnames,table]{xcolor} 
\usepackage[colorlinks=true,
            linkcolor=blue,
            urlcolor=blue,
            citecolor=BrickRed]{hyperref}
            
\usepackage{mathrsfs}  

\usepackage{tabulary}

\usepackage{booktabs}

\usepackage{mathtools}
\usepackage{amssymb}
\usepackage{amsmath}

\usepackage[abs]{overpic}
\usepackage{xcolor,varwidth}   
\definecolor{light-gray}{gray}{0.95}

\usepackage{fancyhdr}

\newcommand{\conferenceheader}{%
  \small\itshape
Proceedings of the 15th International ERCOFTAC Symposium on Engineering Turbulence and Measurements (ETMM-15), Dubrovnik, Croatia.%
}

\fancypagestyle{plain}{
  \fancyhf{}
  \fancyhead[C]{\conferenceheader}
  \fancyfoot[C]{\thepage}

}

\begin{document}
\title{Uncertainty analysis of URANS simulations coupled with an anisotropic pressure fluctuation model
}

\author{Ali Eidi}
\email{a.eidi@tudelft.nl}
\affiliation{Department of Flow Physics \& Technology, Faculty of Aerospace Engineering, TU Delft, 2629 HS Delft, The Netherlands}
\author{Richard P.~Dwight}
\affiliation{Department of Flow Physics \& Technology, Faculty of Aerospace Engineering, TU Delft, 2629 HS Delft, The Netherlands}


\begin{abstract}
Accurate prediction of pressure and velocity fluctuations in turbulent flows is essential for understanding flow-induced vibration and structural fatigue. This study investigates the role of turbulence model parameter uncertainty in such predictions using a combination of global sensitivity analysis, surrogate modeling, and Bayesian inference. The methodology is applied to two fluid-only flow cases: turbulent channel flow and  turbulent annular flow. In the channel flow case, calibrated parameter distributions lead to improved agreement with reference data. In the annular case, limited parameter identifiability is observed, though predictions remain consistent with high-fidelity trends. The results demonstrate both the potential and limitations of model calibration strategies in wall-bounded turbulent flows.
\end{abstract}
\maketitle 
\pagestyle{fancy}
\thispagestyle{fancy}
\section{Introduction}\label{sec:1}
Turbulence-induced vibrations (TIV) in nuclear reactors threaten fuel rod integrity. Axial coolant flow generates pressure fluctuations that cause vibrations, leading to wear, fatigue, and structural damage. Accurate prediction of these effects is critical to reactor safety and component longevity.
Fluid–structure interaction (FSI) simulations are widely used to analyze TIV. While high-fidelity approaches such as Direct Numerical Simulation (DNS) and Large Eddy Simulation (LES) offer detailed turbulence resolution, they are often too expensive for FSI studies. As a practical alternative, Unsteady Reynolds-Averaged Navier–Stokes (URANS) simulations are commonly employed \cite{de2013modal}, though their reliance on turbulence closure models introduces significant uncertainty in pressure fluctuation predictions.

To address this, synthetic turbulence models have been proposed to enhance URANS predictions without resolving all turbulence scales \cite{smirnov2001random}. One example is the pressure fluctuation model, designed to reconstruct missing pressure fluctuations \cite{kottapalli2017numerical}. The anisotropic pressure fluctuation model (AniPFM) extends this idea by incorporating anisotropic velocity fluctuations, improving pressure field predictions in TIV studies \cite{zwijsen2024optimized}. However, AniPFM depends on several tunable parameters, and URANS-AniPFM simulations often fail to match pressure statistics accurately within reasonable parameter ranges \cite{zwijsen2024development}.

Turbulence models based on RANS formulations also introduce uncertainty due to empirical tuning and simplified assumptions \cite{duraisamy2019turbulence}. Case-specific adjustments are typically needed to improve predictive performance. Recent studies have demonstrated the potential of Bayesian methods to quantify parameter variability and enhance predictive robustness in RANS modeling \cite{edeling2014bayesian, edeling2014predictive}.
Given these challenges, sensitivity analysis (SA) and Bayesian calibration have become essential tools to quantify uncertainty and refine turbulence model parameters using high-fidelity data. This study applies such a framework to URANS-AniPFM simulations using the SST $k$-$\omega$  model in a fluid-only configuration. A structured methodology is employed to quantify uncertainties and calibrate model parameters, as outlined in Section 2. Section 3 presents preliminary results and key findings, followed by conclusions and an outlook on further studies in Section 4.

\section{Methodology} \label{sec:2}
\subsection{The $k - \omega$ SST turbulence model}
The Shear Stress Transport (SST) $k-\omega$ model \cite{menter1994two} combines $k-\omega$ for near-wall accuracy with $k-\varepsilon$ for improved free-stream behavior. Its blending function enhances robustness in complex flows, particularly under adverse pressure gradients.
The SST model includes multiple empirical parameters that influence its performance ($\sigma_k$,$\beta_1$, $\beta^*$,$a_1$, and $\kappa$ among others). These parameters are typically determined from canonical experiments and empirical tuning, but remain uncertain when applied to complex flow conditions. A detailed summary of these parameters and their uncertainty limits is provided by \cite{zhang2022uncertainty}.

\subsection{Anisotropic pressure fluctuation model}

The anisotropic pressure fluctuation model (AniPFM) enhances URANS predictions by reconstructing missing pressure fluctuations using synthetic anisotropic velocity fields. Starting from a Reynolds decomposition of the velocity and pressure fields, the pressure fluctuation $p'$ is obtained by solving the Poisson equation:
\begin{equation}
\label{eq:pfm}
\begin{aligned}
\frac{\partial^2 p'}{\partial x_i \partial x_i} &= 
- \rho \bigg[ 
\frac{\partial}{\partial x_i} \left( u'_j \frac{\partial \bar{u}_i}{\partial x_j} \right) \\
& \quad + \frac{\partial^2}{\partial x_i \partial x_j} (u'_i u'_j - \overline{u'_i u'_j}) 
\bigg],
\end{aligned}
\end{equation}
where \( \bar{u}_i \) is the mean velocity and \( u'_i \) the fluctuating velocity component. The accuracy of \( p' \) depends on the quality of the synthetic velocity field.
AniPFM generates fluctuating velocity components \( u'_i \) using a spatial Fourier decomposition, where the dimensionless fluctuations \( w_i(\mathbf{x}, t) \) are given by:
\begin{equation}
w_i(\mathbf{x}, t) = \sqrt{6} \sum_{n=1}^{N} \sqrt{q_n} \, \sigma_{in} \cos\left( \mathbf{k}_n \cdot (\mathbf{x} - \bar{\mathbf{u}} t) + \varphi_n \right),
\end{equation}
with \( q_n \) the mode amplitude, \( \mathbf{k}_n \) the wavenumber vector, \( \sigma_{in} \) the directional cosine, and \( \varphi_n \) a random phase. The amplitudes \( q_n \) are sampled from a modified von Kármán energy spectrum:
\begin{equation}
E(k) = \frac{(k/k_t)^4}{\left[1 + 2.4(k/k_t)^2\right]^{17/6}} \exp\left(-\frac{12 k}{k_\eta}\right),
\end{equation}
where \( k_t \) is the energy-containing peak wavenumber and \( k_\eta \) the dissipation-range cutoff.
To enforce anisotropy, the dimensionless velocity field is transformed using the Cholesky decomposition of the local Reynolds stress tensor \( R_{ij} \):
\begin{equation}
u'_i = a_{ij} w_j, \quad \text{with } a_{ik} a_{kj} = R_{ij},
\end{equation}
ensuring that the synthesized fluctuations reproduce the anisotropic structure predicted by URANS.
This model improves URANS predictions of unsteady pressure fields in wall-bounded flows, particularly in configurations relevant to flow-induced vibrations. A full formulation is available in \cite{zwijsen2024development}.

\subsection{Simulation cases}
Two canonical turbulent flow configurations are used to evaluate the URANS-AniPFM framework: \textit{turbulent channel flow} at multiple friction Reynolds numbers and a more application-relevant \textit{turbulent annular flow} representing rod-bundle geometries.

\subsubsection{Channel flow cases}
The channel flow setup is chosen for its simplicity, statistical homogeneity, and availability of high-fidelity DNS data. Simulations are conducted at four friction Reynolds numbers: $Re_\tau = 180$, 395, 640, and 1020, covering a range of turbulence intensities.
Each domain consists of parallel walls with periodic boundaries in the streamwise ($x$) and spanwise ($z$) directions. A constant body force maintains the target $Re_\tau$. Structured meshes with sufficient $y$-direction resolution ensure grid-independent statistics. 
Simulations use the SST $k$-$\omega$ model with second-order time integration and a CFL limit of 0.5. A two-step approach is employed: steady-state RANS using \texttt{simpleFoam}, followed by unsteady URANS with \texttt{pimpleFoam} and AniPFM. Each run covers 10–15 flow-through times. Time-averaged results are compared against DNS data from \cite{abe2001direct} and others.

\subsubsection{Annular flow case}
To test the framework in a more complex geometry, annular flow between concentric cylinders is simulated at $Re_H = 45{,}000$ with a bulk velocity of $U_{\text{bulk}} = 3.45~\text{m/s}$. The domain has length $L = 6 D_h$, where $D_h$ is the hydraulic diameter. Axial periodicity is imposed, with no-slip walls at the inner and outer cylinders.
The same simulation strategy as in the channel flow is used. The unsteady run spans about 100 flow-through times. Validation is performed against LES data from \cite{norddine2023wall}.

\subsection{Uncertainty framework}

To assess the impact of turbulence model parameter uncertainty on URANS-AniPFM simulations, a four-step uncertainty quantification (UQ) framework is employed: primary sensitivity analysis (SA), surrogate modeling, Bayesian calibration, and secondary SA.
The first step uses a Sobol-based global SA to rank the influence of SST model parameters on the simulation outputs. This allows focusing the calibration on the most significant parameters, reducing computational cost.
A Kriging surrogate is then built to approximate the relationship between selected parameters and quantities of interest (QoIs). To account for structural discrepancies, a Gaussian process-based model inadequacy term is added:
\begin{equation}
\label{eq:kriging}
\hat{y}(\boldsymbol{\theta}) = \mathcal{M}_{\text{Kriging}}(\boldsymbol{\theta}) + \delta(\boldsymbol{\theta}),
\end{equation}
where $\hat{y}(\boldsymbol{\theta})$ is the corrected prediction, $\mathcal{M}_{\text{Kriging}}$ is the Kriging output, and $\delta(\boldsymbol{\theta})$ is a zero-mean Gaussian process (Kennedy and O'Hagan, 2001). This hybrid surrogate supports robust calibration and uncertainty propagation.

Bayesian inference is used to calibrate model parameters with high-fidelity data. Markov Chain Monte Carlo (MCMC) sampling yields posterior parameter distributions and predictive uncertainty bounds for QoIs.
A secondary SA is then performed using the surrogate model to assess the influence of calibrated parameters and verify model robustness.

\section{Results and Discussion} \label{sec:3}
The uncertainty analysis considers nine independent parameters from the SST $k$-$\omega$ model: $\alpha_{k1}$, $\alpha_{k2}$, $\alpha_{\omega1}$, $\alpha_{\omega2}$, $\beta_1$, $\beta_2$, $\beta^*$, $\kappa$, and $a_1$. A first-order Sobol SA is performed to identify the most influential parameters for each case. The sampling matrix is constructed using Saltelli’s method, resulting in 1408 URANS-AniPFM simulations based on $N = 128$ base samples.
From the Sobol results, the most influential four parameters are retained for surrogate modeling. A Kriging model is constructed for each QoI and used for Bayesian calibration and uncertainty propagation. The next subsections present results separately for the channel and annular flow cases.

\subsection{Turbulent Channel Flow (TCF)} 
\label{sec:3.1}
\begin{figure}[h]
    \centering
    \includegraphics[width=0.6\textwidth]{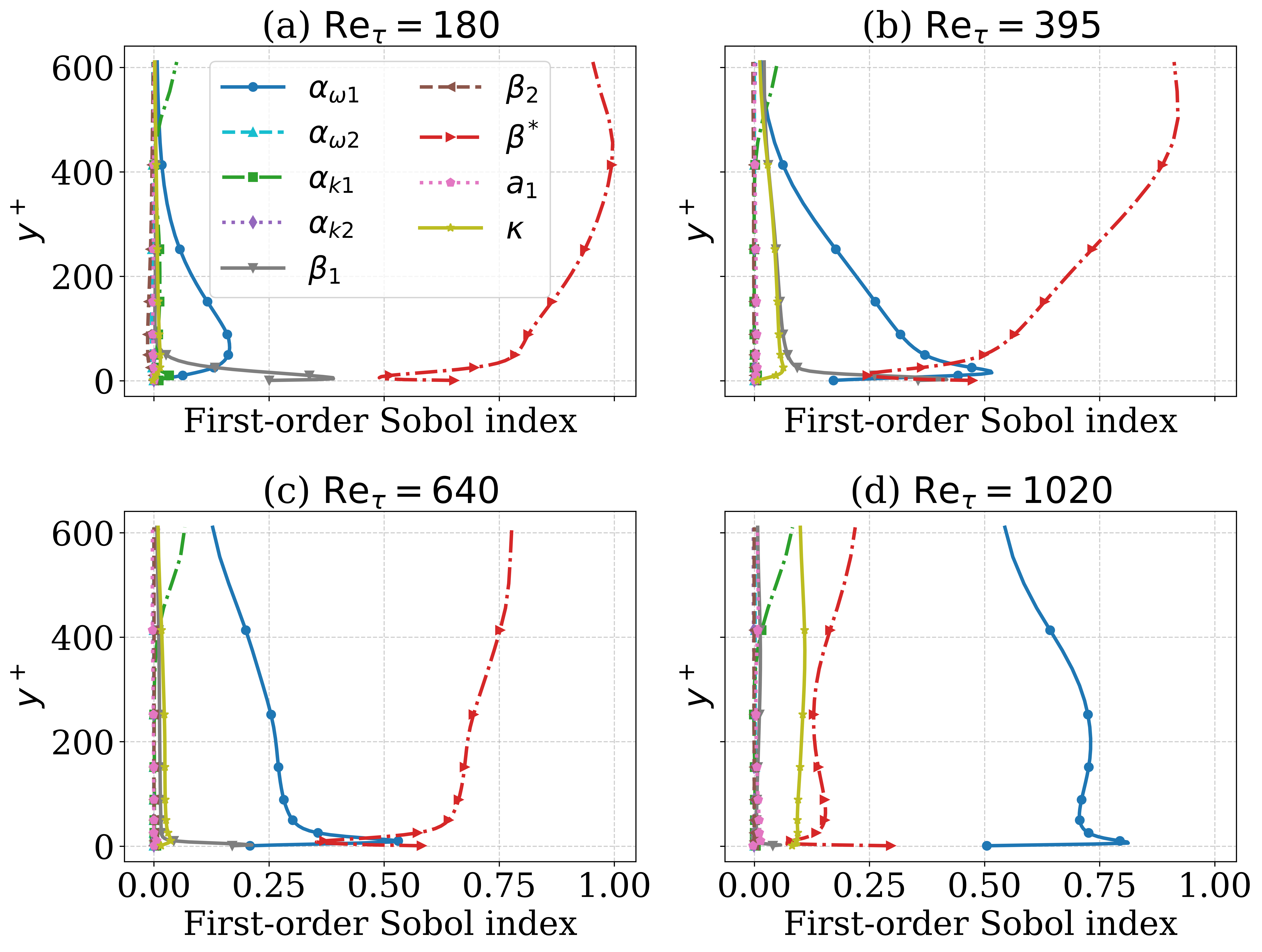} %
    \caption{First-order Sobol sensitivity indices of the SST $k$–$\omega$ turbulence model parameters with respect to the turbulent kinetic energy ($k$) profiles across four friction Reynolds numbers:
(a) $\mathrm{Re}_{\tau} = 180$,
(b) $\mathrm{Re}_{\tau} = 395$,
(c) $\mathrm{Re}_{\tau} = 640$, and
(d) $\mathrm{Re}_{\tau} = 1020$.}
    \label{fig:sobol_k}
\end{figure}

\begin{figure}[h]
    \centering
    \includegraphics[width=0.6\textwidth]{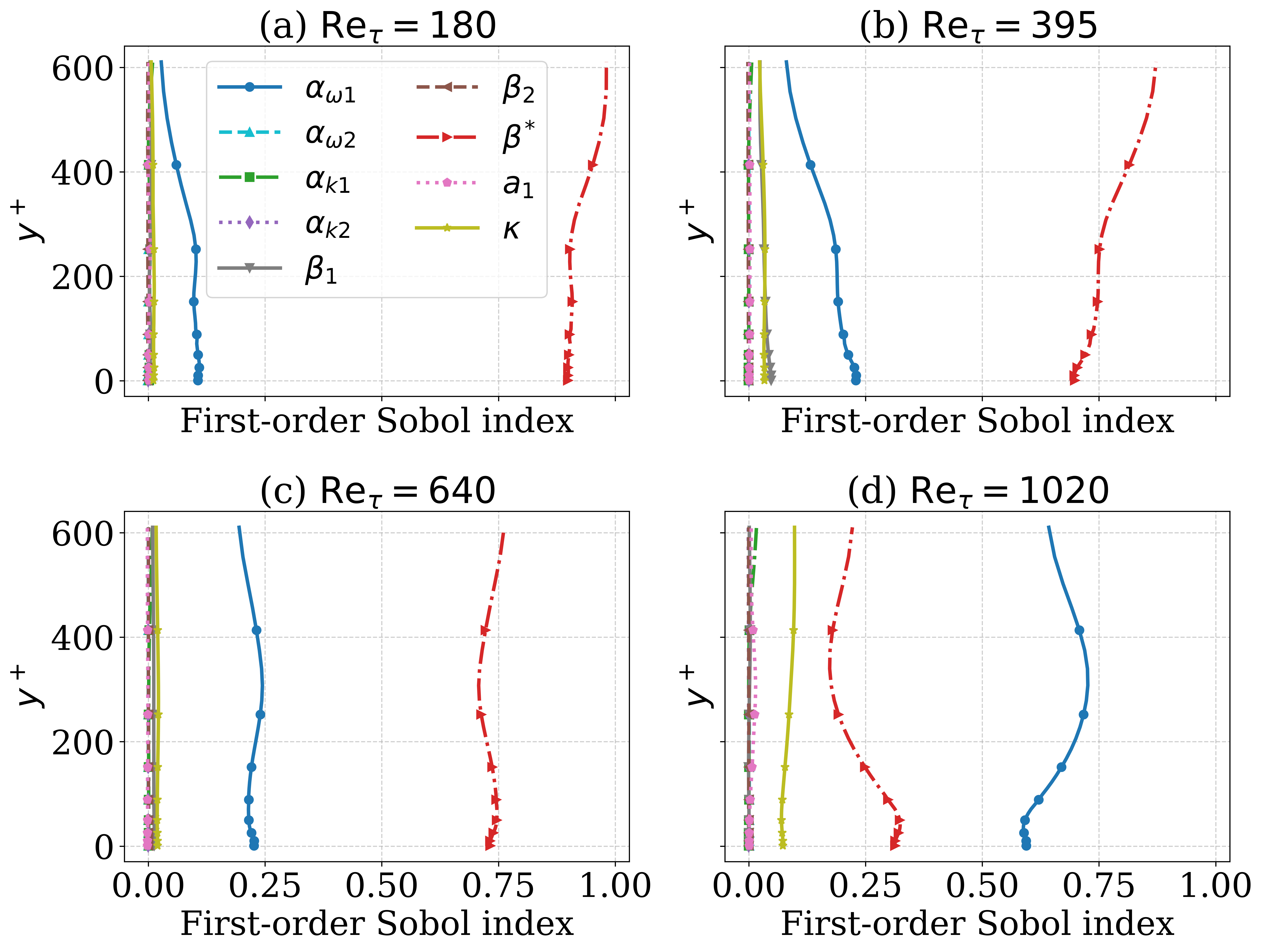} %
    \caption{First-order Sobol sensitivity indices of the SST $k$–$\omega$ turbulence model parameters with respect to the normalized root-mean-square pressure fluctuations ($p_{\mathrm{RMS}}^+$) for four friction Reynolds numbers:
(a) $\mathrm{Re}_{\tau} = 180$,
(b) $\mathrm{Re}_{\tau} = 395$,
(c) $\mathrm{Re}_{\tau} = 640$, and
(d) $\mathrm{Re}_{\tau} = 1020$.}
    \label{fig:sobol_pRMS}
\end{figure}

For the TCF case, a primary SA was performed across four different friction Reynolds numbers ($Re_\tau = 180$, 395, 640, and 1020) to investigate Reynolds number sensitivity and identify influential parameters across flow regimes. Full surrogate-based calibration and UQ analysis were carried out for the $Re_\tau = 640$ case only. This choice was motivated by the availability of high-fidelity DNS data, the focus of prior AniPFM studies on this Reynolds number, and the need to maintain computational feasibility. 
The first-order Sobol sensitivity indices were computed for different QoIs including the turbulent kinetic energy ($k$) and the normalized root-mean-square pressure fluctuations ($p_{\mathrm{RMS}}^+$). The results are presented in Figures~\ref{fig:sobol_k} and~\ref{fig:sobol_pRMS}.

Overall, the parameters $\alpha_{\omega1}$ and $\beta^*$ consistently dominate the sensitivity for both $k$ and $p_{\mathrm{RMS}}^+$ across all cases, especially in the near-wall and outer regions, respectively. The parameters $\beta_1$ and $\kappa$ also show notable influence, though their impact is more localized and Reynolds-number dependent. Similar patterns are observed in the sensitivity of Reynolds stress components (not shown), which follow trends consistent with those of the turbulent kinetic energy.

The sensitivity trends across Reynolds numbers indicate a shift in the dominant parameters as flow complexity increases. For $k$, the influence of $\alpha_{\omega1}$ becomes more pronounced at higher $Re_\tau$, especially in the near-wall region, while $\beta^*$ consistently shows strong influence in the outer region across all cases. The parameter $\beta_1$ retains notable sensitivity in the near-wall region at lower $Re_\tau$, but its impact diminishes as Reynolds number increases. The turbulent kinetic energy also shows increasing sensitivity to $\kappa$ with rising $Re_\tau$, suggesting its growing role in controlling the outer-layer behavior. 

For $p_{\mathrm{RMS}}^+$, similar trends are observed: $\alpha_{\omega1}$ remains consistently influential across all Reynolds numbers and wall-normal positions, while $\beta^*$ maintains a strong contribution, particularly away from the wall. In contrast to the $k$ sensitivity, $\beta_1$ does not exhibit a dominant role near the wall for pressure fluctuations, and $\kappa$ gains modest importance at higher $Re_\tau$.

\begin{figure}[h]
    \centering
    \includegraphics[width=0.6\linewidth]{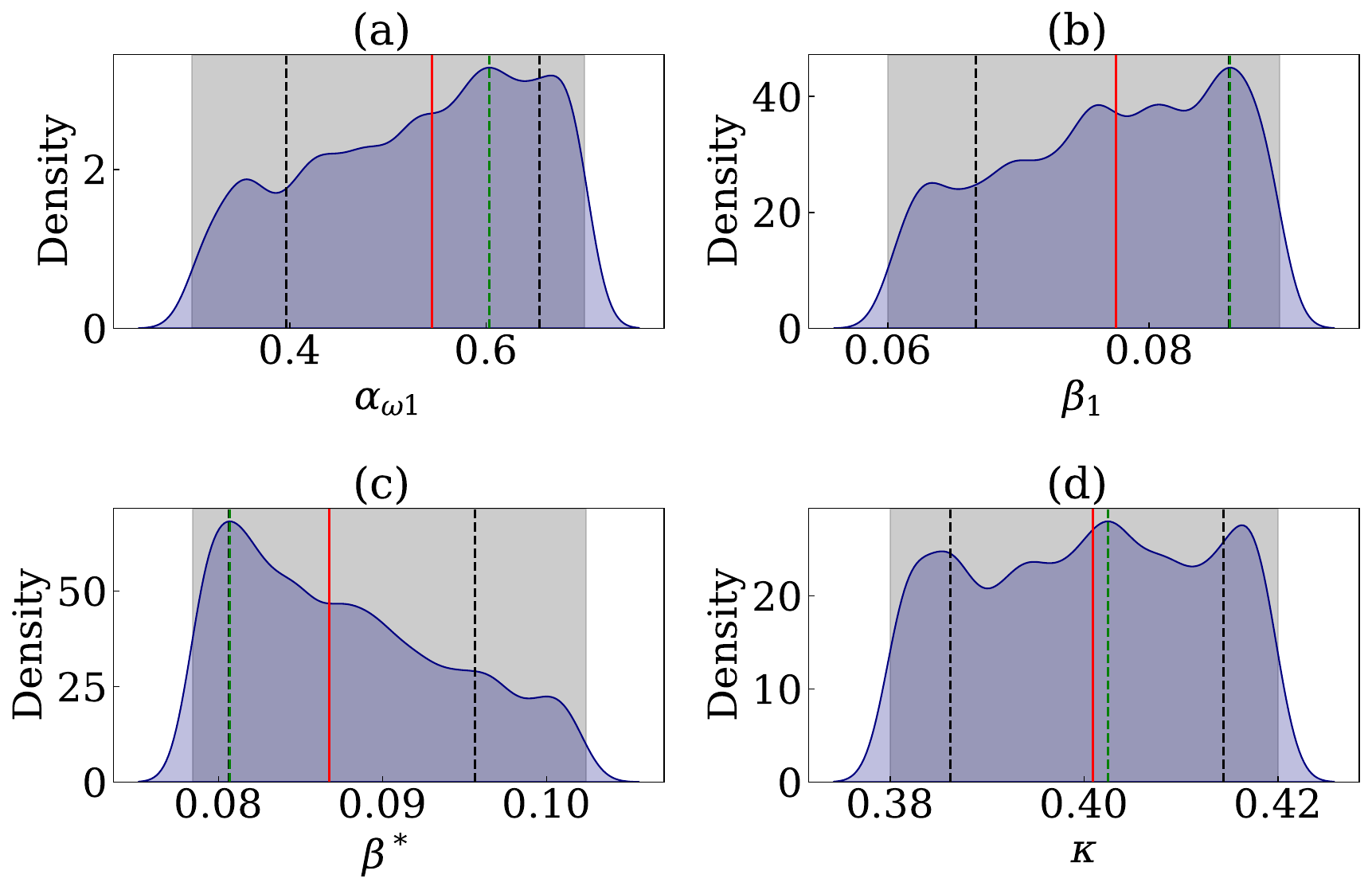}
    \caption{Posterior probability density functions of the four selected SST $k$–$\omega$ model parameters calibrated at $Re_\tau = 640$. Subplots show: (a) $\alpha_{\omega1}$, (b) $\beta_1$, (c) $\beta^*$, and (d) $\kappa$. Dashed vertical black lines denote the 68\% credible intervals; dashed vertical green line indicate the posterior mode. Uniform priors are shown as gray bands.}
    \label{fig:posterior_tcf}
\end{figure}

Based on these observations, the four most influential parameters—$\alpha_{\omega1}$, $\beta_1$, $\beta^*$, and $\kappa$—were selected for surrogate modeling and Bayesian calibration at $Re_\tau = 640$, which serves as the reference case. These parameters collectively capture both near-wall and outer-layer dynamics, while maintaining consistency across multiple Reynolds numbers and QoIs.
Figure~\ref{fig:posterior_tcf} presents the marginal posterior distributions of the four selected turbulence model parameters—$\alpha_{\omega1}$, $\beta_1$, $\beta^*$, and $\kappa$—obtained through Bayesian calibration at $Re_\tau = 640$. All distributions are computed using MCMC sampling and compared against their original uniform priors (gray bands). The 68\% credible intervals (black dashed lines) and mode values (green dashed line) are annotated for each parameter.

The results reveal varying levels of parameter identifiability. The posterior of $\beta_1$  (Figure~\ref{fig:posterior_tcf}b) shows a right-skewed distribution with a clear peak and moderate contraction relative to its prior, suggesting partial informativeness. Both $\alpha_{\omega1}$ and $\beta^*$ (Figures~\ref{fig:posterior_tcf}a \& c) exhibit well-defined, narrow posterior peaks, indicating strong constraint by the data. The mode of $\beta^*$ notably shifts from the prior center, suggesting that the DNS data favors a slightly lower value than originally assumed. Finally, $\kappa$ (Figure~\ref{fig:posterior_tcf}d) also displays a moderately peaked distribution, with minor reduced spread and a mode close to the prior center, reflecting minor calibration sensitivity.

\begin{figure}[h]
    \centering
    \includegraphics[width=0.6\linewidth]{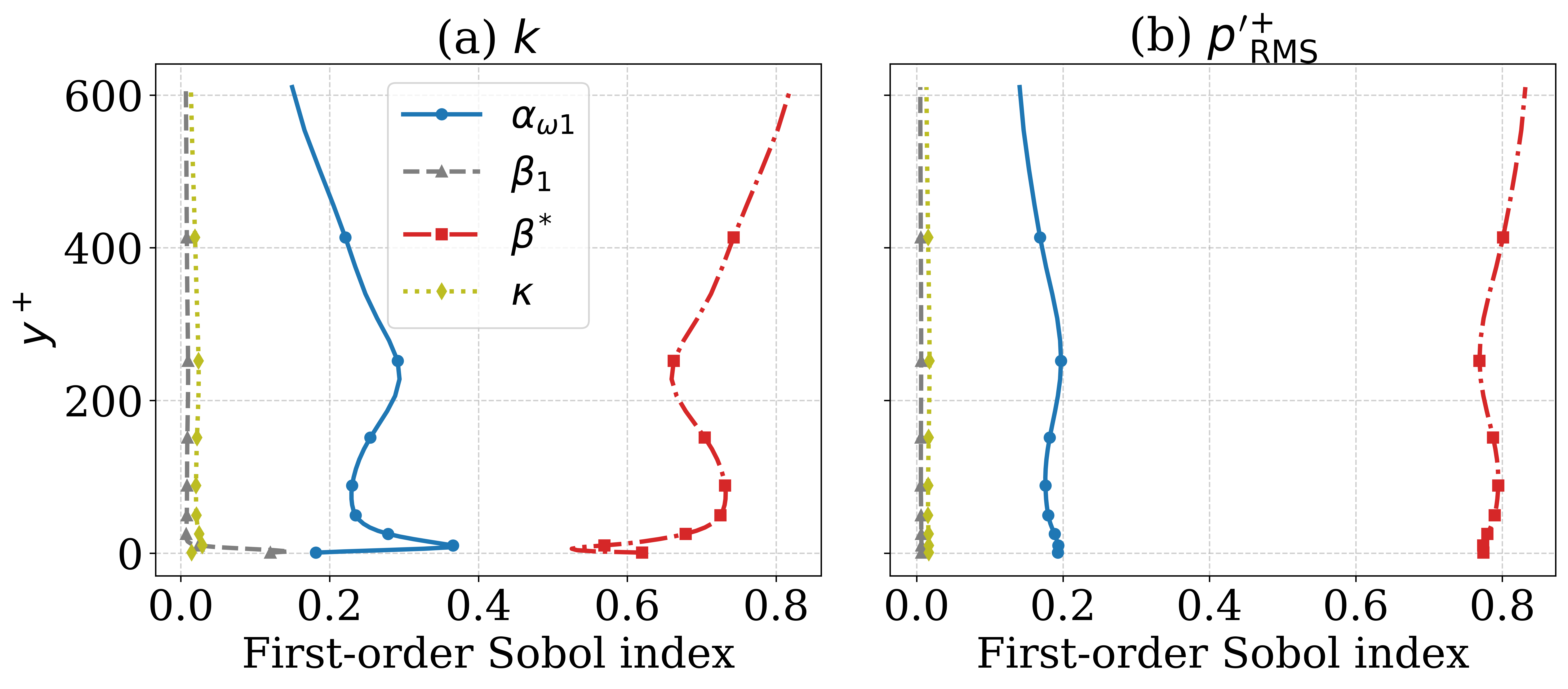}
    \caption{First-order Sobol indices computed from the trained surrogate model with respect to: (a)turbulent kinetic energy ($k$), and (b) normalized pressure fluctuation magnitude ($p'^+_{\mathrm{RMS}}$) at $Re_\tau = 640$.}
    \label{fig:secondary_sobol_tcf}
\end{figure}

Overall, the calibration process effectively narrows the uncertainty in key parameters, particularly for $\beta^*$ and $\alpha_{\omega1}$, which appear to be the most informed by the observed data in this configuration.
To further evaluate the reliability of the surrogate model and to verify the consistency of the calibrated parameter space, a secondary SA was performed using the trained Kriging surrogate. Figure~\ref{fig:secondary_sobol_tcf} presents the first-order Sobol indices of the four selected parameters with respect to $k$ (left) and $p_{\mathrm{RMS}}^+$ (right), using the surrogate model at $Re_\tau = 640$.

The trends observed in this secondary SA closely mirror those obtained from the primary SA using full CFD simulations (Figures~\ref{fig:sobol_k}–\ref{fig:sobol_pRMS}). In particular, $\beta^*$ maintains a dominant influence on both QoIs, while $\alpha_{\omega1}$ plays a major role in the near-wall region of $k$ and retains strong sensitivity for $p_{\mathrm{RMS}}^+$. The influence of $\beta_1$ remains limited to the very near-wall region of $k$, and $\kappa$ contributes modestly in the outer region. This consistency not only confirms the correctness of the surrogate model in reproducing the true model behavior, but also reinforces the validity of the calibrated posterior distributions presented in Figure~\ref{fig:posterior_tcf}.

\begin{figure}[h]
    \centering
    \includegraphics[width=0.6\linewidth]{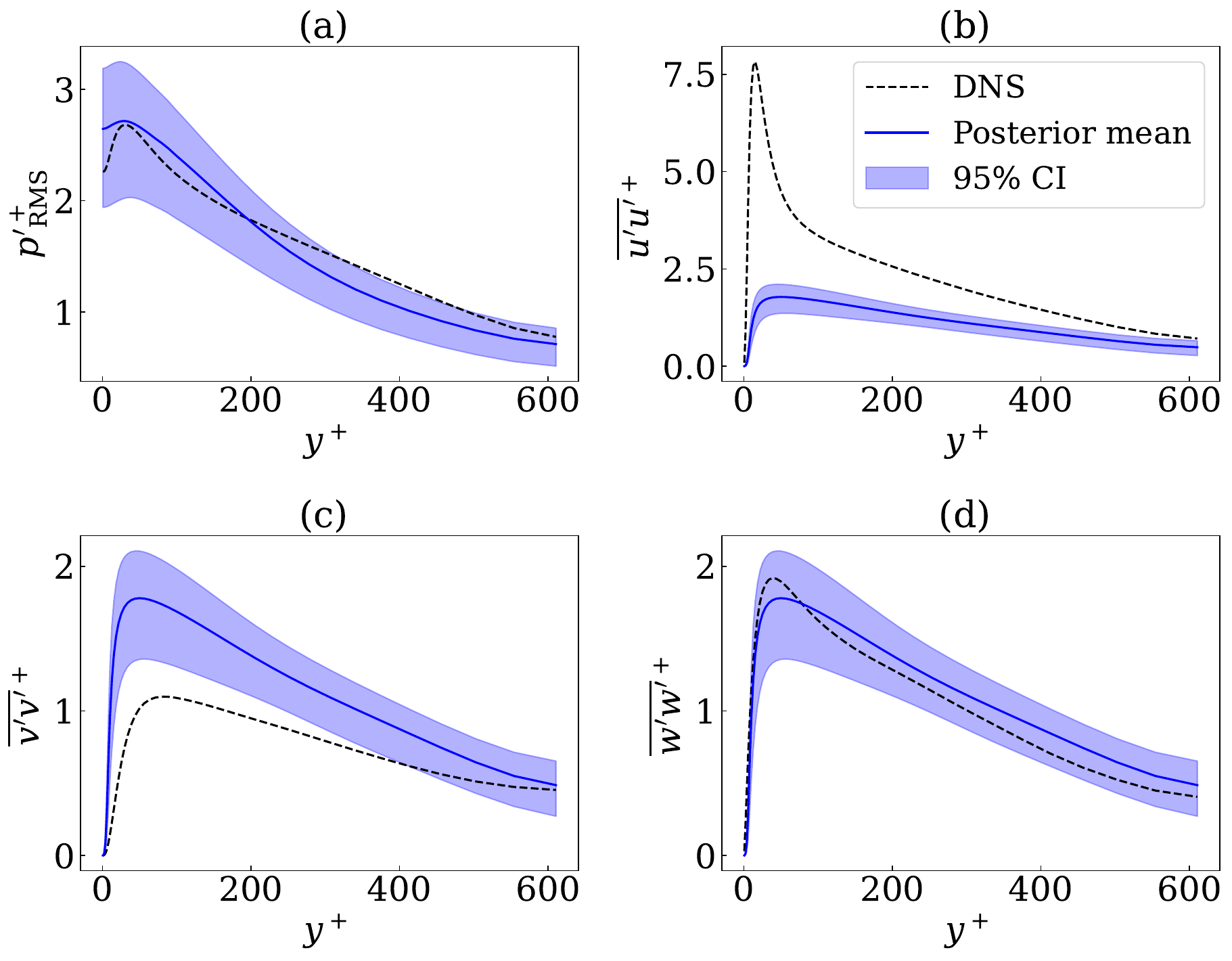}
    \caption{Predictive posterior distributions of key flow quantities at $Re_\tau = 640$: (a) pressure RMS $p'^+_{\mathrm{RMS}}$, (b) streamwise Reynolds stress $\overline{u'u'}^+$, (c) wall-normal Reynolds stress $\overline{v'v'}^+$, and (d) spanwise Reynolds stress $\overline{w'w'}^+$.}
    \label{fig:posterior_prediction_tcf}
\end{figure}

The close agreement between primary and secondary SA demonstrates that the surrogate model captures the underlying physics-driven parametric dependencies and can be confidently used for uncertainty propagation and parameter prioritization.
Figure~\ref{fig:posterior_prediction_tcf} presents the posterior predictive distributions for four key QoIs at $Re_\tau = 640$: ($p'^+_{\mathrm{RMS}}$) and normalized Reynolds stress components $\overline{u'u'}^+$, $\overline{v'v'}^+$, and $\overline{w'w'}^+$. The profiles show the posterior mean and 95\% credible intervals, compared against DNS reference data.

The predictions for $p'^+_{\mathrm{RMS}}$ show good agreement with DNS data across the wall-normal direction, with the DNS curve lying mostly within the credible interval. This is expected since $p'^+_{\mathrm{RMS}}$ was one of the calibration targets. The predictions for $\overline{w'w'}^+$ also follow the DNS profile reasonably well, especially in the outer region, although small deviations exist near the wall.
In contrast, the model exhibits noticeable discrepancies in the near-wall behavior of $\overline{u'u'}^+$ and $\overline{v'v'}^+$. The posterior mean underpredicts the peak of $\overline{u'u'}^+$, and $\overline{v'v'}^+$ is not well captured close to the wall. These deviations may reflect limitations of the URANS-AniPFM framework in accurately reconstructing anisotropic Reynolds stress components in strongly inhomogeneous regions, or insufficient information in the calibration targets to constrain these quantities.

Despite these shortcomings, the fact that the DNS profiles mostly fall within the 95\% predictive intervals for all four quantities indicates that the calibrated model captures the overall uncertainty structure. The results suggest that while the UQ framework improves prediction and quantifies uncertainty effectively for pressure fluctuations and some Reynolds stress components, further refinement—potentially through multi-output calibration or improved surrogate fidelity—may be needed for full accuracy in all Reynolds stress components.


\subsection{Turbulent Annular Flow (TAF)} \label{sec:3.2}
\begin{figure}[h]
    \centering
    \includegraphics[width=0.6\linewidth]{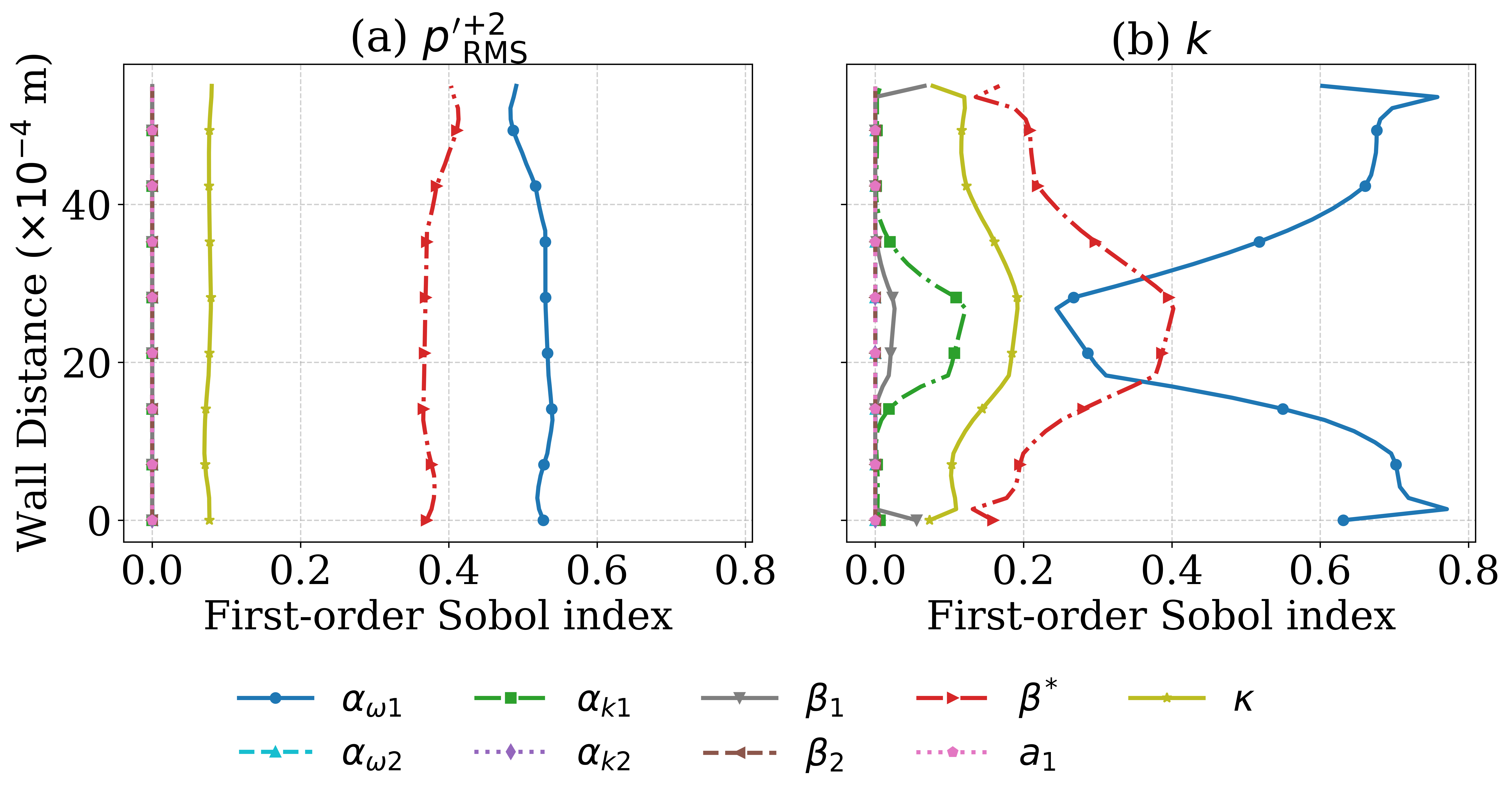}
    \caption{First-order Sobol indices for the annular flow case at $Re_H = 45{,}000$ with respect to: (a) normalized pressure fluctuation magnitude ($p'^{+2}_{\mathrm{RMS}}$), and (b) turbulent kinetic energy ($k$).}
    \label{fig:sobol_taf}
\end{figure}

For the annular flow case, the full uncertainty quantification pipeline was applied at $Re_H = 45{,}000$. Figure~\ref{fig:sobol_taf} presents the first-order Sobol indices computed for pressure fluctuations ($p'^{+2}_{\mathrm{RMS}}$) and turbulent kinetic energy ($k$), with wall-normal profiles plotted radially from the inner cylinder. These results form the basis for parameter selection and calibration in the following steps.

The results reveal that $\alpha_{\omega1}$ and $\beta^*$ are consistently the most influential parameters for both QoIs. In particular, $\alpha_{\omega1}$ exhibits dominant sensitivity in the inner-wall region of $k$, while $\beta^*$ gains increasing importance farther from the wall. These trends are consistent with observations from the TCF configuration and highlight the robustness of these parameters across different geometries.
For $p'^{+2}_{\mathrm{RMS}}$, the dominance of $\alpha_{\omega1}$ and $\beta^*$ is even more pronounced, with other parameters such as $\kappa$, $\alpha_{k1}$, and $\beta_1$ showing only marginal contributions. The remaining parameters have negligible impact, indicating that most of the model response variance in both $k$ and $p'^{+2}_{\mathrm{RMS}}$ is captured by a reduced subset of parameters.
These findings justify the selection of $\alpha_{\omega1}$, $\alpha_{k1}$, $\beta^*$, and $\kappa$ for surrogate modeling and Bayesian calibration in the TAF case.

\begin{figure}[h]
    \centering
    \includegraphics[width=0.6\linewidth]{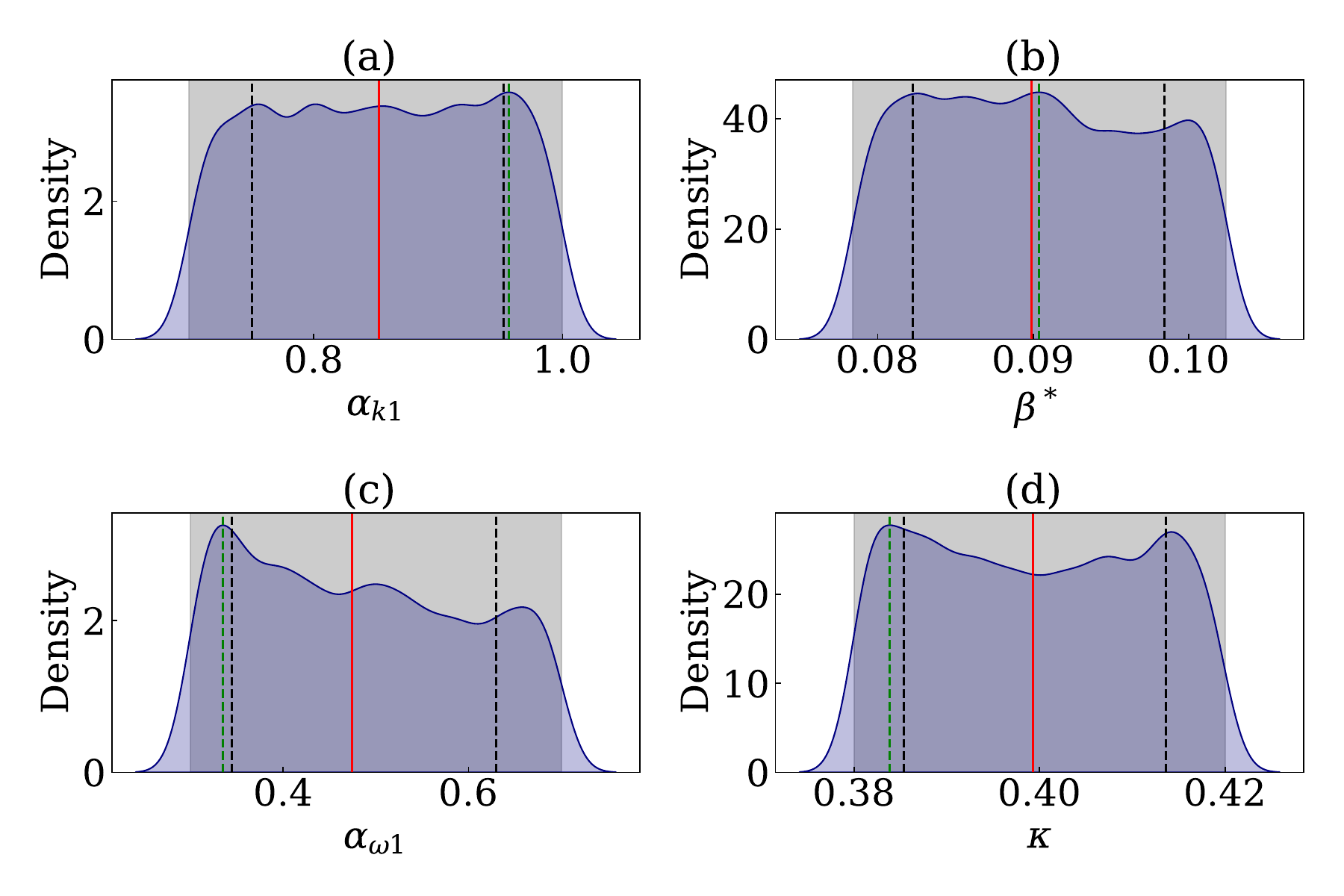}
    \caption{Posterior distributions of the selected SST $k$–$\omega$ model parameters for the annular flow case at $Re_H = 45{,}000$: (a) $\alpha_{k1}$, (b) $\beta^*$, (c) $\alpha_{\omega1}$, and (d) $\kappa$. Dashed vertical black denote the 68\% credible intervals; dashed vertical green line indicate the posterior mode. Uniform priors are shown as gray bands.}
    \label{fig:posterior_taf}
\end{figure}

Figure~\ref{fig:posterior_taf} shows the marginal posterior distributions of the four selected turbulence model parameters—$\alpha_{\omega1}$, $\beta_1$, $\beta^*$, and $\kappa$—obtained through Bayesian calibration of the annular flow case. The posterior modes and 68\% credible intervals are annotated on each subplot.
Unlike the channel flow case, the TAF results reveal limited parameter identifiability. The posterior distributions remain broad and in some cases nearly uniform, particularly for $\alpha_{k1}$ and $\beta^*$, indicating that the available LES data do not strongly constrain these parameters within their prior ranges. Only $\alpha_{\omega1}$ and $\kappa$ exhibit moderate concentration around specific values, but even these retain considerable uncertainty.
These results suggest that either the QoIs used for calibration do not sufficiently inform the full parameter space, or that the model's sensitivity to some parameters is weaker in the TAF geometry compared to the TCF case. 

\begin{figure}[h]
    \centering
    \includegraphics[width=0.6\linewidth]{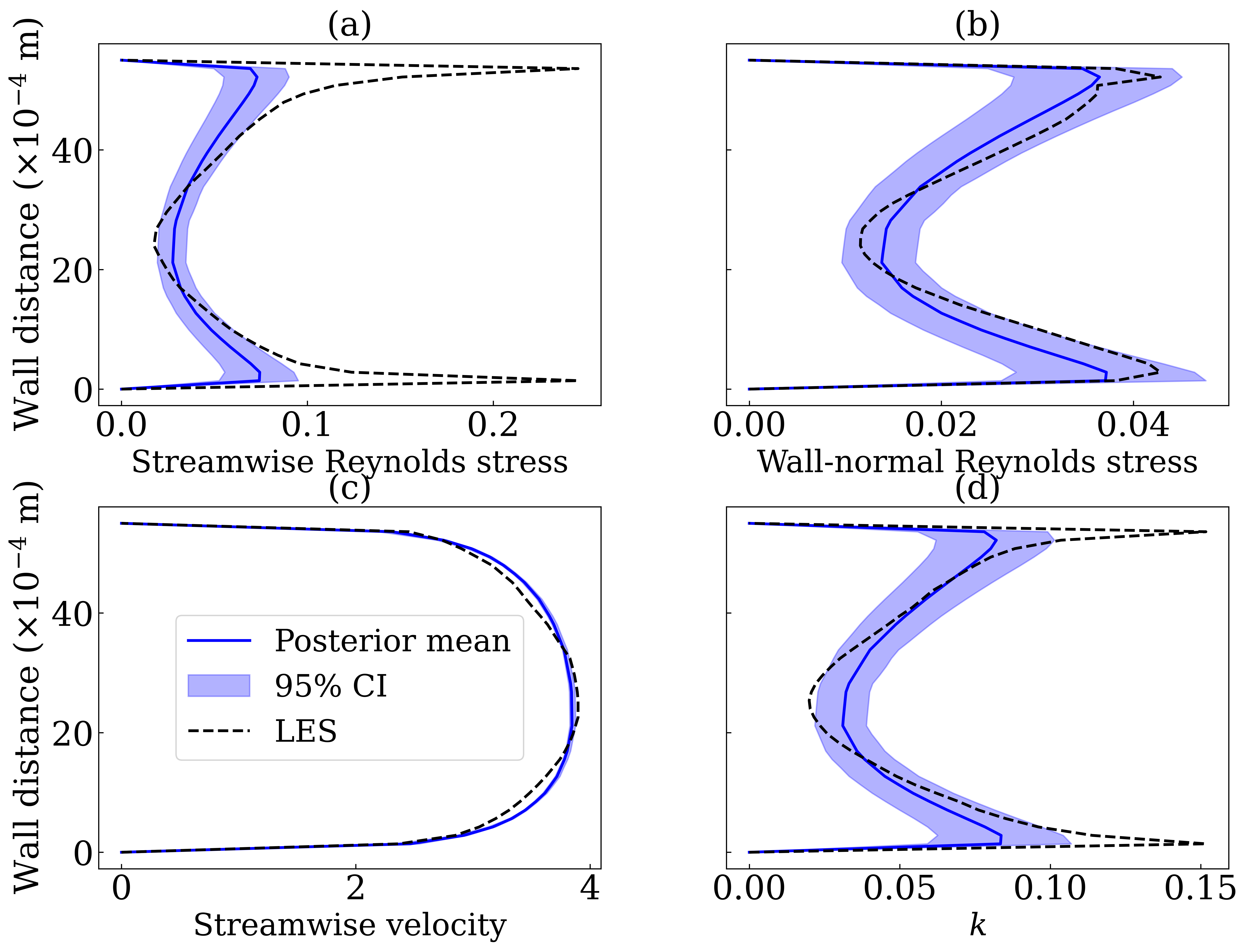}
    \caption{Posterior predictive profiles for the annular flow case at $Re_H = 45{,}000$: (a) streamwise Reynolds stress $\overline{u'u'}$, (b) wall-normal Reynolds stress $\overline{v'v'}$, (c) streamwise mean velocity $\bar{u}$, and (d) turbulent kinetic energy $k$.}
    \label{fig:posterior_prediction_taf}
\end{figure}

Figure~\ref{fig:posterior_prediction_taf} presents the posterior predictive distributions of four key flow quantities for the TAF case at $Re_H = 45{,}000$: streamwise Reynolds stress ($\overline{u'u'}$), wall-normal Reynolds stress ($\overline{v'v'}$), streamwise velocity ($\bar{u}$), and turbulent kinetic energy ($k$). Each plot shows the posterior mean and 95\% credible interval, compared against LES reference data.

Despite this, the predictive profiles agree reasonably well with the LES results—especially for mean velocity—and the credible intervals successfully bound the high-fidelity data. These results suggest that the prior variability already captures much of the model uncertainty. The findings emphasize the need for more informative calibration targets or extended QoIs to better constrain model parameters in complex geometries.
A secondary SA was also conducted using the trained surrogate model and yielded results consistent with the primary SA, but is omitted here for brevity.

\section{Conclusions} \label{sec:4}
This study applied a structured uncertainty quantification framework to wall-bounded turbulent flows, combining SA, surrogate modeling, and Bayesian calibration. Two flow configurations were considered: TCF and TAF.
In the TCF case, SA identified key turbulence model parameters, and Bayesian inference led to well-informed posteriors and improved agreement with reference data. In contrast, calibration for the TAF case resulted in limited parameter constraint, though the predicted profiles still captured key trends.
These findings highlight the importance of parameter identifiability and observable selection in achieving reliable model calibration. Since URANS is coupled with AniPFM, the results also suggest that calibrating turbulence model parameters alone may be insufficient, as the AniPFM model introduces its own uncertainty. In future work, the influential parameters identified in this study will be reused for surrogate-based analysis in more complex TIV-related configurations, where full SA would be computationally prohibitive.

\section*{Acknowledgment}

A.E. acknowledges support from the GO-VIKING project (Grant No. 101060826) under the Euratom Research and Training Programme. The authors also acknowledge use of DelftBlue computational resources provided by the Delft High Performance Computing Centre.

\section*{References}
\bibliographystyle{ieeetr} 
\bibliography{main}
\end{document}